\documentstyle[12pt,aasms4]{article}

\slugcomment{Accepted for the publication in the {\it Astrophysical Journal}}

\def\ros{{\it ROSAT}}
\def\iras{{\it IRAS}}
\newcommand{\oq}{$1\over4$}
\newcommand{\tq}{$3\over4$}
\newcommand{\hi}{H{\small I}}
\newcommand{\hii}{H{\small II}}
\newcommand{\NH}{\mbox{$N_{\rm H}$}}

\def\la{\mathrel{\hbox{\rlap{\hbox{\lower4pt\hbox{$\sim$}}}\hbox{$<$}}}}
\def\ga{\mathrel{\hbox{\rlap{\hbox{\lower4pt\hbox{$\sim$}}}\hbox{$>$}}}}

\lefthead{Park, Finley, \& Dame}
\righthead{X-ray Shadows}

\begin{document}
\title{EVIDENCE FOR THE GALACTIC X--RAY BULGE II}
\author{\bf Sangwook Park and John P. Finley}
\affil{Department of Physics,  Purdue University \\
 1396 Physics Building, West Lafayette, IN. 47907 \\
Electronic Mail: (parksan,finley)@physics.purdue.edu}
\author{\bf and \\
           T. M. Dame}
\affil{Harvard-Smithsonian Center for Astrophysics, 60 Garden Street,
Cambridge, MA. 02138 \\
Electronic Mail: tdame@cfa.harvard.edu}


\begin{abstract}
A mosaic of 5 \ros~PSPC pointed observations in the Galactic plane
($l\sim25^{\circ}$) reveals X-ray shadows 
in the $0.5-2.0$~keV band cast by distant molecular clouds. The 
observed on-cloud and off-cloud X-ray fluxes indicate 
that $\sim15$\% and $\sim37$\% of the diffuse X-ray background in this 
direction in the \tq~keV and 1.5~keV bands, respectively, originates 
behind the molecular gas which is located at $\sim$3 kpc from the Sun.  
The implication of the derived background X-ray flux beyond the absorbing 
molecular cloud is consistent with, and lends further support to 
recent observations of a Galactic X-ray bulge. 

\end{abstract}


\section {\label {sec:intro} INTRODUCTION}

The origin and nature of the diffuse X-ray background (DXB) in the 
0.5 -- 2.0 keV band has been a puzzle for the last three decades. 
In the Galactic plane, study of the 0.5 -- 2.0 keV band DXB is 
complicated by the presence of discrete Galactic emission features 
(e.g., supernova remnants) which must be accounted for in any
investigation of the diffuse component. Despite the difficulties 
it can be safely assumed that the 0.5 -- 2.0 keV DXB in the plane is 
Galactic in origin since extragalactic emission is
completely absorbed and contributions due
to unresolved stellar sources have been shown to be small
(\cite{ss90}; \cite{w92}; \cite{os92}). 
Provided that the contribution from Galactic discrete emission features 
can be effectively removed,
the Galactic plane is, then, a useful laboratory for the study of 
the Galactic component of the DXB. 
One way to study the 3-dimensional spatial structure 
of this Galactic component is to search for
``shadows'' cast by absorbing molecular 
gas at various distances and directions. The \ros~XRT/PSPC is 
an ideal tool for this type of study with its large field of view
(2$^{\circ}$ diameter), good angular resolution (25$''$ on-axis), and 
soft X-ray bandpass (0.1 -- 2.4 keV).
Previous studies at high Galactic latitudes utilizing the \ros~PSPC 
(e.g., \cite{sel91}; \cite{bm91}; \cite{smv93}; \cite{wy95})
have demonstrated the utility of the ``shadowing'' technique, which 
is enhanced when the absorbing cloud is at a known distance.

Recently, X-ray shadows in the DXB cast by distant molecular
gas in the Galactic plane have been detected in the direction of 
{\it l} $\sim$ 10$^{\circ}$ with a mosaic of \ros~PSPC pointed 
observations (\cite{pel97}, PFSD hereafter). PFSD detected a
shadowing depth in both the \tq~keV and 1.5 keV bands of $\sim$43\% and the 
derived background intensity (beyond the absorbing molecular gas) is 
more than an order of magnitude brighter
than the nominal high latitude intensity. This 
striking result implies the existence of an X-ray emitting Galactic 
bulge. This observation is consistent with analysis of \ros~all-sky survey 
data in the general direction of the Galactic center (\cite{sel97}) 
where excess emission observed at high Galactic latitudes was
extrapolated down to the Galactic plane.
PFSD determined that the de-absorbed background spectrum of the 
tentative Galactic X-ray bulge is consistent with an $\sim$10$^{6.7}$ 
K thermal plasma, also in good agreement with the all-sky survey
results.
Here we report the detection of X-ray shadows cast by distant molecular 
clouds in the direction of {\it l} $\sim$ 25$^{\circ}$ in the Galactic 
plane. 

There is probably a much more complicated mixture of X-ray emission and
absorption along the line of sight toward {\it l} $\sim$
25$^{\circ}$ than toward the {\it l} $\sim$ 10$^{\circ}$
region studied by PFSD.  Toward {\it l} $\sim$ 10$^{\circ}$, 
the line of sight passes through the molecular ring and then through a 
long section of the Galactic center region where the density of both 
atomic and molecular gas is very low and the X-ray emission, presumably, 
originates (the X-ray bulge). Most of the neutral and molecular gas 
therefore lies in front 
of most of the X-ray emission, and clouds are seen clearly in silhouette.  
Toward {\it l} $\sim$ 25$^{\circ}$, molecular clouds are more widely
distributed along the line of sight, which passes nearly tangent to the 
inner edge of the molecular ring and through a shorter chord of the X-ray
bulge.  

The data used for this study are described in \S\ref{sec:data}.  
The analysis and results are presented in \S\ref{sec:results} and the
implications are discussed in \S\ref{sec:discussion}. 

\section{\label{sec:data} DATA}

Previous studies of the DXB using X-ray shadows cast by
nearby molecular clouds have utilized the anticorrelation between
the X-ray intensity and the \iras~100 $\mu$m intensity (e.g.,
\cite{bm91}; \cite{bm94}; \cite{smv93}; \cite{wy95}; \cite{ksv97}). 
The \iras~100 $\mu$m data are a reasonable gas tracer 
and the utility of these data has been demonstrated with 
high-latitude molecular clouds with {\it known} distances 
(e.g., Draco Cloud at $\sim$300 pc, \cite{bm91}, \cite{sel91}; 
MBM12 at $\sim$65 pc, \cite{smv93}) 
or nearby low-latitude dense molecular clouds (e.g., Coalsack at $\sim$200 
pc from the Sun, \cite{bm94}). 
However, to search for 0.5 -- 2.0 keV band X-ray shadows 
cast by ``distant'' ($>$1 kpc) molecular clouds in the Galactic plane, 
the \iras~100 $\mu$m data are of limited use, since there are generally many 
clouds overlapped along the line of sight and with no velocity
information there is no way to determine their individual distances.
The \iras~100 $\mu$m data can also be confused by
emission from \hii~regions in the plane. These difficulties with
the \iras~100 $\mu$m data can be overcome by studying 
the X-ray anticorrelations with CO
spectral line data, a standard tracer of interstellar molecular gas. 
Although CO data also have their own complications such as the conversion
between CO intensity and H column density, they clearly provide better 
information regarding absorbing molecular clouds in the plane.
This work is thus based upon information derived from such data.

\subsection{\label{subsec:X-ray} X-ray Data}

The soft X-ray data utilized in this study consist of 
5 \ros~(\cite{tru92}) PSPC pointed observations (see 
Table~\ref{tbl:observations}). Data from all 5 pointings 
were obtained through the High Energy Astrophysics Science Archive
Research Center (HEASARC) \ros~public archive. The R4, R5, R6, and R7
band data (cf., \cite{smbm94} for band definitions) are utilized in 
the study of the 0.5 -- 2.0 keV band DXB in the Galactic plane. 
The R1L and R2 band emission, comprising a \oq~keV band, is assumed 
to have a completely 
different origin; i.e., the Local Hot Bubble (LHB), a region of
$\sim$10$^{6}$ K plasma surrounding the Sun with an average extent of 
$\sim$100 pc (\cite{cr87}). This emission is 
thus not considered in this study. 

Since non-cosmic contamination must be carefully handled 
when studying the DXB, all identified non-cosmic contributions 
to the counting rate must be excluded by time selection or modeled 
and subtracted from the data. 
We modeled and subtracted the particle background, scattered solar X-ray 
background, and long-term enhancements from the data.
A detailed description of the non-cosmic background subtraction methods 
can be found in the literature (\cite{smbm94}). 
We also eliminated contamination due to short-term enhancements including
auroral X-rays, solar flares, and enhanced charged-particle rates encountered
near the South Atlantic Anomaly and particle belts by excluding all
observation time intervals which display anomalous peaks in their light curves.
For these 5 pointings, the mean contribution of the modeled non-cosmic 
contamination to the counts is $\sim$13\% in the 0.5 -- 2.0 keV band.
We removed the detected point sources and extended discrete emission 
features (e.g., SNRs) from the data prior to the DXB analysis.
Unresolved point sources are expected to provide only a minor
contribution to the surface brightness.

After all identified non-cosmic contamination, detected point sources,
and discrete extended emission features are removed, the individual PSPC 
pointings are merged into large-area mosaics in two bands: the \tq~keV 
band (R4+R5) and the 1.5 keV band (R6+R7). The determination 
of the relative offsets in the zero level between the individual
observations is performed by comparing the average count rates in the
overlapping regions between all pairs of observations. The contribution
from this correction is typically small ($\sim$2\% and $\sim$3\% of the
total counts in the \tq~keV band and the 1.5 keV band, respectively) 
for the entire field of view.  The software for this
task (\cite{s94}) was provided by the US \ros~Science Data Center (USRSDC) 
at NASA/GSFC. 

The final \ros~PSPC mosaics of the $l~\sim25^\circ$ direction of the
Galactic plane
are displayed in Figure~\ref{fig:xr}a (the \tq~keV band) and 
Figure~\ref{fig:xr}b (the 1.5 keV band). 
For the purpose of display, the data were smoothed to 7.5$'$ resolution in
order to match the resolution of the CO data used.
The circles in Figure~\ref{fig:xr}a
indicate the 10 regions where detected bright point sources and discrete 
extended emission features have been removed. 
Combined as a mosaic, the average exposure for this field is $\sim9$~ks 
yielding $\sim16$\% (1.5 keV band) and $\sim20$\% (\tq~keV band) 
statistical errors for individual 5$'$ pixels.

\subsection{\label{subsec:CO} CO Data}

The CO data used in the present study were obtained with the CfA 1.2 m
telescope as part of a large on-going survey of the first Galactic
quadrant (\cite{dt94}). At 115 GHz, the frequency of the J = 1 -- 0
rotational transition, the telescope has a beamwidth of 8.4$'$ (FWHM) 
and its 256-channel spectrometer provides a velocity resolution of 0.65
km s$^{-1}$ and total bandwidth of 166 km s$^{-1}$. Observations were
spaced roughly every beamwidth (7.5$'$) on a Galactic grid, and the rms
sensitivity of 0.14 K (T$_{mb}$) was more than adequate for the present
purpose.

\section{\label{sec:results} ANALYSIS \& RESULTS}

Figure~\ref{fig:xrco} compares the 1.5 keV band X-ray image with contours
of CO intensity integrated over 3 different ranges of velocity. 
For the clarity of the presentation of the CO intensity variation, 
gray-scale CO images for the same 3 velocity intervals are displayed in 
Figure~\ref{fig:co}. The overall anticorrelation 
between the X-ray and the CO intensities, as presented in Figure~\ref{fig:xrco},
reveals several X-ray shadows. These X-ray shadows are indicated by
solid-lined rectangles in Figure~\ref{fig:xr}b and labeled as A, B, C, 
and D. 

Shadow A at {\it l,b} $\sim$ 
23.5$^{\circ}$,0$^{\circ}$ appears to arise mainly from a molecular 
cloud at a velocity of  $\sim$ 54 km s$^{-1}$ (Figure~\ref{fig:xrco}b)
for which the near kinematic distance is $\sim$3.8 kpc.
Despite the clear shadowing feature in this region, a detailed
analysis of Shadow A is unwarranted due to the difficulty in
extracting a meaningful off-cloud surface brightness (see below for more
discussion).

It is notable that a much more intense CO feature at 
{\it l,b} $\sim$ 23.4$^{\circ}$,--0.35$^{\circ}$ in the V = 55 -- 130 
km s$^{-1}$ map (Figure~\ref{fig:xrco}c) does not produce any noticeable
X-ray shadow. This intense CO feature is believed to be a large
star-forming cloud at the far kinematic distance (d $\sim$ 13 kpc; 
\cite{del86}). Assuming that the bulk of the detected diffuse X-ray 
emission is originating from within or foreground to the Galactic 
X-ray bulge, no detected shadow by this CO enhancement is reasonable
since this cloud should be located behind the 0.5 -- 2.0 keV band X-ray 
emitting region.

Shadow B at {\it l,b} $\sim$ 24.6$^{\circ}$,0$^{\circ}$ 
seems to be associated with the molecular clouds in the velocity 
interval V = 30 -- 55 km s$^{-1}$ 
(Figure~\ref{fig:xrco}b) which corresponds to a near kinematic distance of
$\sim$3 kpc. The molecular clouds in the velocity interval 55 -- 130
km s$^{-1}$, however, seem to contribute substantially to this shadow 
as well (Figure~\ref{fig:xrco}c). This complexity may arise from the
existence of numerous molecular clouds along this line of sight due to
the Galactic structure in this direction. As a matter of fact, the
velocity-integrated CO intensity in the region of Shadow B  ({\it
l,b} $\sim$ 24$^{\circ}$ 
-- 25$^{\circ}$, 0$^{\circ}$)  is the strongest
in the entire Galaxy outside of the Galactic center (\cite{del87}).

The origin of Shadow C is unclear.  The CO intensity in this region is
relatively strong in the velocity range of 30 -- 130 km s$^{-1}$ 
(Figure~\ref{fig:xrco}b and Figure~\ref{fig:xrco}c), but 
Figure~\ref{fig:xrco}a suggests that
lower-velocity gas may also contribute to the shadow, which would be
reasonable as at least some of the gas is likely to be nearby.  

Shadow D at {\it l,b} $\sim$ 25.5$^{\circ}$,0.7$^{\circ}$ is the
most remarkable.  There is a clear anticorrelation between this feature and a
well-defined molecular cloud at V = 46 km s$^{-1}$ (Figure~\ref{fig:xrco}b).  
With a near kinematic distance of 3.3 kpc (Figure~\ref{fig:dvsv}), 
the molecular cloud extends fairly high above the Galactic plane 
(z$\sim$70 pc at 3.3 kpc, z$\sim250$ pc at the far kinematic distance,
12.2 kpc) where there is much 
less foreground or background
molecular gas, but still a bright X-ray background.  Shadow D is thus the most
appropriate for a detailed quantitative analysis of the X-ray absorption. 
Applying a CO-H$_{2}$ conversion factor $\frac{N(H_2)}{W_{CO}}$ = 1.9 $\times$
10$^{20}$ cm$^{-2}$ (K km s$^{-1}$)$^{-1}$ (\cite{sm96}),
the H$_2$ column density corresponding to the mean W$_{CO}$ within the Shadow D
region is $\sim$5.4 $\times$ 10$^{21}$ cm$^{-2}$ for the CO cloud at
$\sim$3 kpc.
This column density implies that the absorbing molecular cloud at 3 kpc 
in this region is relatively optically thick for the 0.5 --
2.0 keV band X-rays since one optical depth is $\sim$2.7 $\times$
10$^{21}$ cm$^{-2}$ and $\sim$4.0 $\times$ 10$^{21}$ cm$^{-2}$ in the
\tq~keV band and the 1.5 keV band, respectively, assuming the theoretical
absorption cross section of Morrison \& McCammon (1983) and a
10$^{6.7}$ K thermal plasma with a foreground \NH$\sim$9 $\times$
10$^{21}$ cm$^{-2}$ (to correct for the hardening of the X-ray bulge spectrum
by the foreground ISM).  We thus directly
extract the on-cloud and off-cloud X-ray intensities in order to
estimate the foreground and the background (behind the molecular cloud)
fractions of the observed X-ray intensity rather than fitting 
the standard two-component absorption model (e.g., \cite{smv93}).

The average on-cloud 1.5 keV band X-ray intensity for Shadow D is  
121$\pm$2 $\times$ 10$^{-6}$ counts s$^{-1}$ arcmin$^{-2}$. 
Due to the presence of many molecular clouds in the plane 
($|$b$|$ $\la$ 0.5$^{\circ}$) at all velocities along 
the line of sight, extracting off-cloud intensities within 
$|${\it b}$|$ $\sim$ 0.5$^{\circ}$ 
of the Galactic plane is not feasible near {\it l} $\sim$
25$^{\circ}$. The off-cloud 1.5 keV band X-ray
intensity is thus estimated in the areas of the three 
dotted-rectangular regions labeled 
a -- c in Figure~\ref{fig:xr}b, where I$_{CO}<7$~K km
s$^{-1}$ (i.e., H$_{2}$ column density $<$ 1.3 $\times$ 10$^{21}$ 
cm$^{-2}$) in
Figure~\ref{fig:xrco}b. The off-cloud intensities are 203$\pm$6,
198$\pm$8, and 177$\pm$6 $\times$ 10$^{-6}$ counts s$^{-1}$
arcmin$^{-2}$ for regions a, b, and c, respectively. 
Combining data from the 
off-cloud X-ray regions yields an off-cloud intensity of 192$\pm$4 
$\times$ 10$^{-6}$ counts s$^{-1}$ arcmin$^{-2}$. 

In the \tq~keV band, the on-cloud and the off-cloud X-ray
intensities are extracted from the same regions as in
the 1.5 keV band. The average on-cloud \tq~keV band X-ray
intensity is 87$\pm$2 $\times$ 10$^{-6}$ counts s$^{-1}$ arcmin$^{-2}$. 
The off-cloud intensities are 113$\pm$4, 96$\pm$6, and 97$\pm$4
$\times$ 10$^{-6}$ counts s$^{-1}$ arcmin$^{-2}$ for region a, b, and c,
respectively. Combining the data from the off-cloud regions yields an 
off-cloud X-ray intensity of 102$\pm$3 $\times$
10$^{-6}$ counts s$^{-1}$ arcmin$^{-2}$. The observed on-cloud and
off-cloud X-ray intensities in the \tq~keV band and the 1.5 keV band are 
summarized in Table~\ref{tbl:fluxes}. The errors for the on-cloud and
off-cloud X-ray intensities in both \tq~and 1.5 keV bands are formal
statistical uncertainties and any systematic errors such as uncertainty due
to the selection of the extraction regions will increase the 
uncertainties presented here.

The average on-cloud to off-cloud X-ray intensity ratio in the 1.5 keV
band is $0.63\pm0.02$, which implies an $\sim$37\% contribution to the
observed diffuse X-ray background from beyond the molecular
cloud. Using \hi~and CO spectra toward the off-cloud regions, 
the foreground column density is estimated to be 
$\sim$7.4 $\times$ 10$^{21}$ \hi~cm$^{-2}$, $\sim$11.1 $\times$
10$^{21}$ \hi~cm$^{-2}$, and $\sim$12.7 $\times$ 10$^{21}$
\hi~cm$^{-2}$ for regions a, b, and c, respectively.
The \hi~spectrum from the 21 cm survey of Hartmann \& Burton (1997)
(angular resolution of 0.5$^{\circ}$) was utilized for the estimation.
To derive total gas column densities along the line of sight,
21 cm and CO emissions were partitioned between the near and
far kinematic distances based on assumed constant scale heights of
220 pc (\cite{loc84}) and 120 pc (\cite{bel88}), respectively.
The average foreground column density for the off-cloud regions is 
10.4$\times$ 10$^{21}$ \hi~cm$^{-2}$ and 
corresponds to a foreground absorption
optical depth in the 1.5 keV band X-rays of $\tau\sim2.6$.
After deabsorption, this implies a
background intensity of 956$\pm$25 $\times$ 10$^{-6}$ counts s$^{-1}$ 
arcmin$^{-2}$ even without any correction for by possible absorption
by material beyond the cloud but foreground to the X-ray emitting
region. This value is, on average, 
about a factor of 7 higher than the nominal high-latitude intensity
($\sim130\times10^{-6}$~counts~s$^{-1}$~arcmin$^{-2}$). 

The average on-cloud to off-cloud X-ray intensity ratio in the \tq~keV
band is $0.85\pm0.03$, which implies that $\sim$15\% of the observed X-ray
emission is from behind the molecular cloud. The foreground absorption
optical depth in the \tq~keV band X-rays is $\tau\sim3.9$ and the
estimated background intensity is 
741$\pm$28 $\times$ 10$^{-6}$ counts s$^{-1}$ arcmin$^{-2}$. 
The derived average background intensity in this energy
band is also higher (by a factor of 6) than the nominal high-latitude
intensity. 

\section{\label{sec:discussion} DISCUSSION}

The derived average background X-ray intensities in both the 1.5 keV and 
\tq~keV band are larger than the nominal high-latitude intensity
by a factor of 6 -- 7, indicating an extensive region
of X-ray emitting Galactic gas beyond 
the absorbing molecular cloud located at $\sim$3 kpc. 
The implied DXB emission is consistent with the existence of an
emission region near the Galactic center as posited by PFSD in
connection with X-ray shadows observed in the direction of 
{\it l} $\sim$ 10$^{\circ}$ and a Galactic X-ray bulge suggested by 
Snowden et al. (1997) based on \ros~all-sky survey data. The derived 
background X-ray intensity (assuming no further absorption behind the
molecular cloud) in the {\it l} $\sim$ 25$^{\circ}$ 
direction, however, appears to be lower than that of PFSD
($\sim$65\% of PFSD in the 1.5 keV band and $\sim$30\% of PFSD 
in the \tq~keV band
on average). The lower values of the derived background 
intensity are not surprising and are likely related to the 
spatial structure of the Galactic X-ray bulge in the plane in that 
the $l~\sim~25^{\circ}$ direction is closer to the ``edge'' of the 
Galactic X-ray bulge and therefore the path length through the emission
region will be shorter. Considering the Galactic molecular gas distribution 
(i.e., the molecular ring; see below for more discussion), the path length
difference through the X-ray bulge between the directions of PFSD ({\it
l} $\sim$ 10$^{\circ}$) and the present paper ($l~\sim~25^{\circ}$) is
expected to produce $<$50\% background X-ray intensity in the 
$l~\sim~25^{\circ}$ direction (Figure~\ref{fig:geo}).

The lower intensity of the DXB behind the absorbing 
molecular cloud and the relatively shallow shadowing depths compared 
with PFSD may also be due to an increase in the interstellar 
absorption beyond 3 kpc. This speculation is 
reasonable when the morphology of the prospective Galactic X-ray bulge 
in the \ros~all-sky survey maps (the highly-enhanced X-ray emission 
feature around the Galactic center region, cf. \cite{sel97}) and the
Galactic structure of molecular gas (see \S\ref{sec:intro}) are
considered. If the weaker shadows and the lower DXB intensity 
are due to additional interstellar absorption beyond 3 kpc, the 
additional absorbing background ISM can be estimated 
using a simple absorption model. Assuming a negligible interstellar 
absorption of the X-ray bulge emission beyond the shadow-casting molecular 
clouds in the {\it l} $\sim$ 10$^{\circ}$ direction, 
the difference in the derived background X-ray intensities between 
{\it l} $\sim$ 10$^{\circ}$ and {\it l} $\sim$ 25$^{\circ}$ (assuming
the average shadowing depths) requires an additional Galactic
interstellar absorption of \NH~$\sim$ 3.2 $\times$ 
10$^{21}$ cm$^{-2}$ beyond 3 kpc in the {\it l} $\sim$ 25$^{\circ}$ 
direction. 
Assuming a foreground column density of 9.6 $\times$ 10$^{21}$ cm$^{-2}$
in the direction of Shadow D as estimated from the \hi~and CO spectra,
the derived total \NH~(foreground + background of the shadow-casting 
molecular cloud) in the {\it l} $\sim$ 25$^{\circ}$ 
direction is $\sim$13 $\times$ 10$^{21}$ cm$^{-2}$. 
For comparison, an analysis of the CO and
\hi~spectra indicates that there should be \NH~of 6.7 $\times$ 10$^{21}$ 
cm$^{-2}$ between 4 kpc and 8 kpc toward the center of the Shadow D 
({\it l,b} = 25.5$^{\circ}$, 0.75$^{\circ}$). The corresponding
total \NH~ is then $\sim$16 $\times$ 10$^{21}$ cm$^{-2}$. 
From the CO and \hi~surveys (\cite{hb97}) the total 
\NH~is estimated to be 18 -- 45 $\times$
10$^{21}$ cm$^{-2}$ in this direction, which is consistent with the
derived total \NH.  

As determined by the Galactic CO gas distribution, the inner edge of the
molecular ring is located at {\it l} $\sim$ 25$^{\circ}$ in the plane.
The Galactic X-ray 
bulge then appears confined by the molecular ring, at least in the
plane, since it extends to only {\it l} $\sim$ 25$^{\circ}$ 
in the first quadrant of the plane (\cite{sel95}). By comparison,
in the {\it l} $\sim$ 10$^{\circ}$ direction, the implied background
fraction of the DXB likely originates within the void 
which is interior to the molecular ring.
The velocity intervals of the absorbing gas identified as shadows in 
the DXB (i.e., 
15 -- 30 km s$^{-1}$ at {\it l} $\sim$ 10$^{\circ}$, PFSD, and 30 -- 
55 km s$^{-1}$ at {\it l} $\sim$ 25$^{\circ}$, the present paper) 
in fact track the molecular ring structure (\cite{del87}). 
The coincidence of the distance scales for the absorbing molecular gas 
($\sim$3 kpc from the Sun) in the directions of both PFSD and the present 
paper with the location of the molecular ring ($\sim$3 -- 4 kpc from 
the Sun at its closest approach) indicates that all of these X-ray 
shadows are cast by 
molecular clouds associated with the Galactic molecular ring. 

The 1.5 keV to \tq~keV band hardness ratio for the average derived 
background emission is 1.29$\pm$0.06 (with no further absorption).
This hardness ratio is higher than that of PFSD (0.59$\pm$0.16). 
The harder spectrum is consistent with the existence of
additional interstellar absorption beyond 3 kpc. 
Assuming the additional absorption is \NH~$\sim$ 3.2 -- 6.7 $\times$ 
10$^{21}$ cm$^{-2}$, the average 1.5 keV to \tq~keV hardness ratio of 
the ``de-absorbed'' spectrum is then $\sim$0.58 -- 0.88, which is closer
to the value of PFSD. These hardness ratios suggest a
plasma temperature of the derived background X-ray emission of T $\ga$
10$^{6.7}$ K.

\section{\label{sec:summary} SUMMARY AND CONCLUSIONS}

We have presented an analysis of X-ray shadows in the 0.5 -- 2.0
keV band diffuse X-ray background in the Galactic plane cast by dense 
molecular clouds in the direction of {\it l} $\sim$ 25$^{\circ}$ 
and discussed the implications of the results. The average on-cloud 
and off-cloud
X-ray intensities imply that $\sim$37\% of the observed 1.5 keV band 
and $\sim$15\% of the \tq~keV band diffuse X-rays originate from beyond
2.6 -- 3.9 optical depths. The average derived background emission beyond
$\sim$3 kpc is a factor of 5 -- 7 higher than the nominal
high-latitude intensity, even with no additional absorption, and confirms 
the existence of the Galactic X-ray bulge (PFSD; \cite{sel97}). 
The intensity of the derived background 
emission, however, appears to be lower than that of PFSD. This lower intensity 
is likely a result of the structure of the Galaxy where the 
$l~\sim~25^{\circ}$ direction of the plane (the present paper) 
is near the {\it edge} of the Galactic 
X-ray bulge while the $l~\sim~10^{\circ}$ direction (PFSD) is 
more toward the {\it center} of the X-ray bulge. The morphology of the 
highly enhanced diffuse X-ray emission around the Galactic center region 
of the \ros~all-sky survey maps (\cite{sel95}) support this suggested angular 
extension of the Galactic X-ray bulge. The distance scales to the 
absorbing clouds producing the X-ray
shadows and the angular extension of the Galactic X-ray bulge may
imply that the 3-dimensional structure of the Galactic X-ray bulge is 
associated with, and likely confined by, the Galactic molecular ring, 
at least in the Galactic plane.

The Galactic X-ray bulge most likely contributes a substantial fraction
of the observed soft X-ray background in the 
Galactic plane. Detailed spatial and spectral studies 
are needed to understand its 
nature and origin. The question of the nature of the soft X-ray 
background making up the foreground flux to the absorbing clouds (which 
may also be responsible 
for the non-zero flux in the Galactic anti-center direction) still 
remains unanswered. Follow-up observations of the detected 
X-ray shadows presented in this paper and PFSD (on-cloud and off-cloud 
regions) and a search for more X-ray shadows cast by molecular
clouds at various distances and directions will help unveil the multi-component 
structure of the soft X-ray background in the Galactic plane.

\acknowledgments

{The authors thank S. Snowden for helpful comments and suggestions for
the clarity of the presentations in this paper.
This research has made use of data obtained through the High Energy
Astrophysics Science Archive Research Center Online Service, provided by
the NASA-Goddard Space Flight Center and was supported in part by NASA
grants NAG 5-2492, NAG 5-3426, and the Purdue Research Foundation.}

\clearpage

\figcaption[]
{X-ray images of the Galactic plane near $l~\sim25^\circ$.
These cover an area of $\sim10$~deg$^{2}$ between $l = 22.5^\circ$ 
and $l=26.5^\circ$ at $\left|b\right|\leq1.5^\circ$. 
Panels (a) and (b) display the \tq~keV and the 1.5~keV X-ray images, 
respectively, from the mosaic of the 5 \ros~PSPC pointed observations in 
Table~1.  The circles indicate the regions where the
detected bright point sources and discrete extended emission
features are removed.  For the purpose of display, the data have been 
smoothed using a gaussian with $\sigma$ = 7.5$'$ (FWHM) to match the 
resolution of the CO data.
In (b) the X-ray shadows A, B, C, and D are
indicated by solid-lined rectangles.
The off-cloud regions a, b, and c are indicated by dotted rectangles.
\label{fig:xr}}

\figcaption[]{The $^{12}$CO (J = 1 -- 0) intensity in three 
different velocity intervals overlaid on the image of the 1.5 keV 
band X-rays. The CO contour interval is 7 K~km~s$^{-1}$, starting at 
7 K~km~s$^{-1}$. The contour at 7 K~km~s$^{-1}$ corresponds to 
$\sim$0.6 optical depths in the $3\over4$ keV band and $\sim$0.4 
optical depths in the 1.5 keV band.  The CO velocity integration range 
is given above each plot. A small portion of the X-ray map at $l 
< 23^\circ$ is excluded because comparable quality CO data was 
not available there.
\label{fig:xrco}}

\figcaption[]{The gray-scale CO image with the overlaid contours. 
The three velocity intervals and the contour levels are the same as in
Figure 2.
\label{fig:co}}

\figcaption[]{Near kinematic distance vs. radial velocity in the
direction of {\it l,b} = 24.5$^{\circ}$, 0$^{\circ}$, based on the
rotation curve of Burton (1992).
\label{fig:dvsv}
}

\figcaption[]{A schematic diagram for the simplistic geometry of the 
Galactic X-ray bulge and the molecular ring in the plane. 
Region A represents the void which is filled with the X-ray emitting
gas (Galactic X-ray bulge). Region B is the molecular ring. The
{\it l} $\sim$ 25$^{\circ}$ direction skims the inner edge of the
molecular ring while the {\it l} $\sim$ 10$^{\circ}$ direction passes
through the X-ray bulge.
\label{fig:geo}}

\clearpage

\begin{deluxetable}{lcccc}
\footnotesize
\tablecaption{List of \ros~PSPC Observations.
              \label{tbl:observations}}
\tablewidth{0pt}
\tablehead{
\colhead{Observation ID} & \colhead{$l$} & \colhead{$b$}
& \colhead{Exposure (ks)} & \colhead{Date}
}
\startdata
RP400287N00 & 23.34 & +0.18 & 7.80 & $3-9$ Apr 1993    \nl
RP500008    & 23.53 & +0.32 & 3.67 & 30 Mar 1991   \nl
RP500012    & 24.66 & +0.61 & 4.26 & 3 Apr 1991       \nl
RP400288N00 & 25.09 & +0.55 & 8.60 & 9 Apr 1993 \nl
WP500204N00 & 25.52 & +0.22 & 16.84 & $7-8$ Apr 1993   \nl
\enddata

\end{deluxetable}

\clearpage

\begin{deluxetable}{ccccc}
\centering
\footnotesize
\tablecaption{Shadow D X-ray Fluxes.
\label{tbl:fluxes}}
\tablewidth{0pt}
\tablehead{
\colhead{Band} & \colhead{On-Cloud\tablenotemark{a}} &
 \colhead{Off-Cloud\tablenotemark{a}} & \colhead{Ratio} &
\colhead{Off-cloud \NH\tablenotemark{b}}
}
\startdata
\tq~keV\ & $87\pm2$  & $102\pm3$  & $0.85\pm0.03$ &
10.4$^{+2.3}_{-3.0}$\nl
1.5 keV & $121\pm2$  & $192\pm4$ & $0.63\pm0.02$ &
10.4$^{+2.3}_{-3.0}$\nl
\enddata

\tablenotetext{a}{On-cloud and off-cloud fluxes in units 
of $10^{-6}$ counts s$^{-1}$ arcmin$^{-2}$. The on-cloud flux is averaged
over rectangle D in Figure~\ref{fig:xr}b and the off-cloud flux is
averaged over rectangles a, b, and c in Figure~\ref{fig:xr}b. The errors 
are the formal statistical uncertainty.}
\tablenotetext{b}{The average foreground column density of the 
off-cloud regions a, b, and c in units of 10$^{21}$ cm$^{-2}$.}

\end{deluxetable}

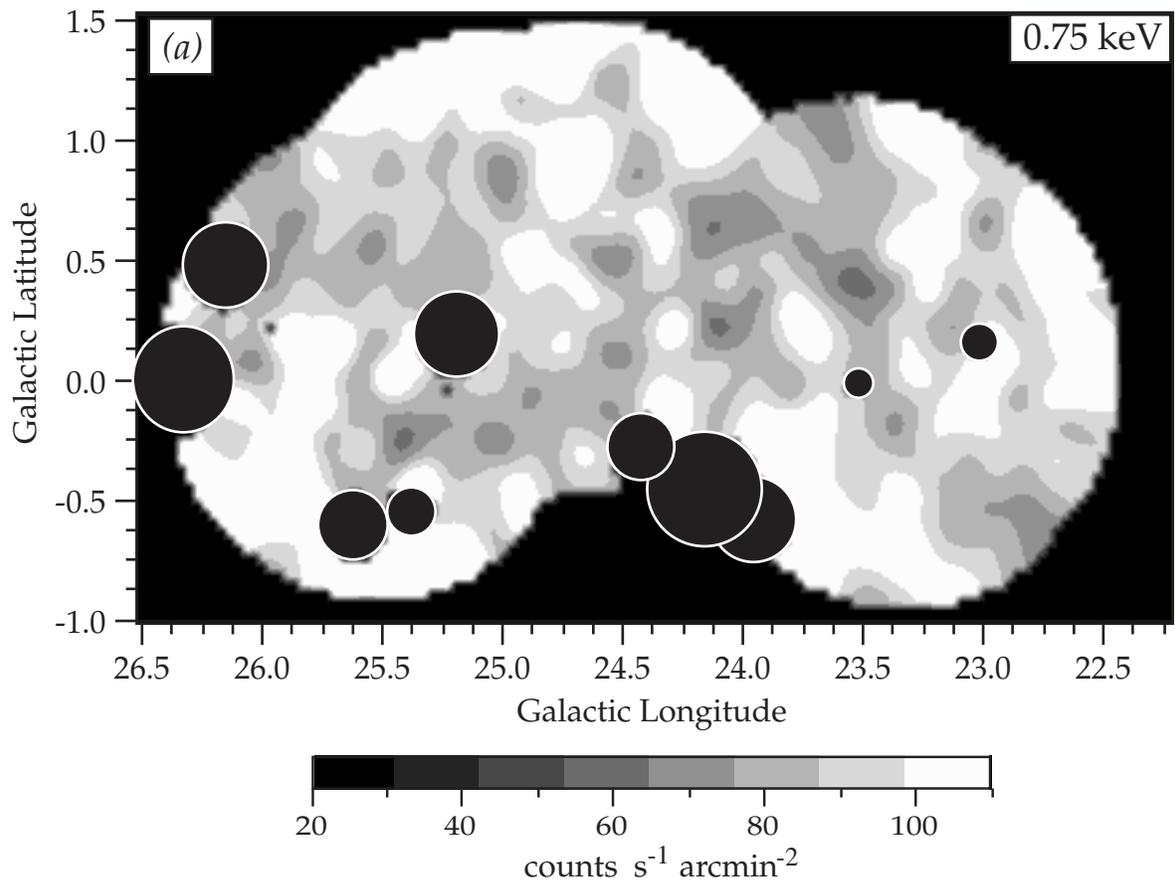
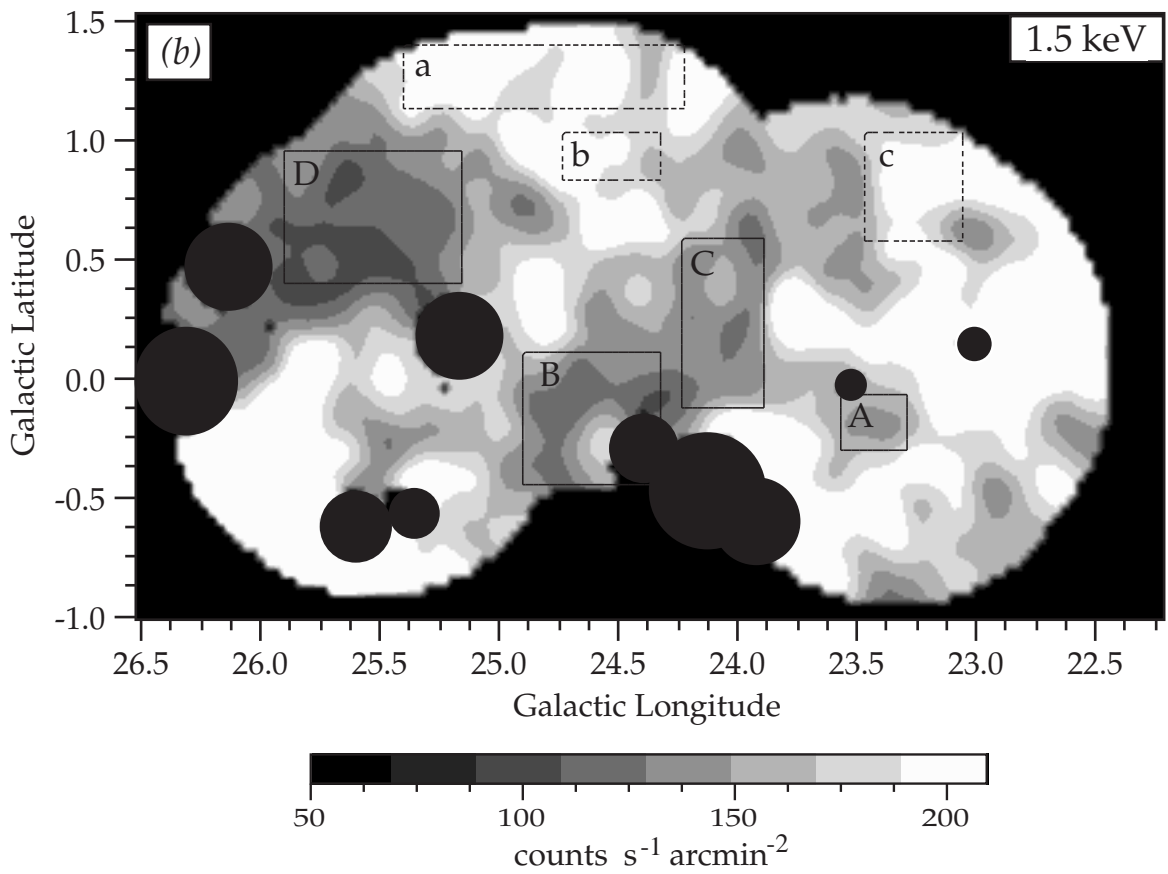
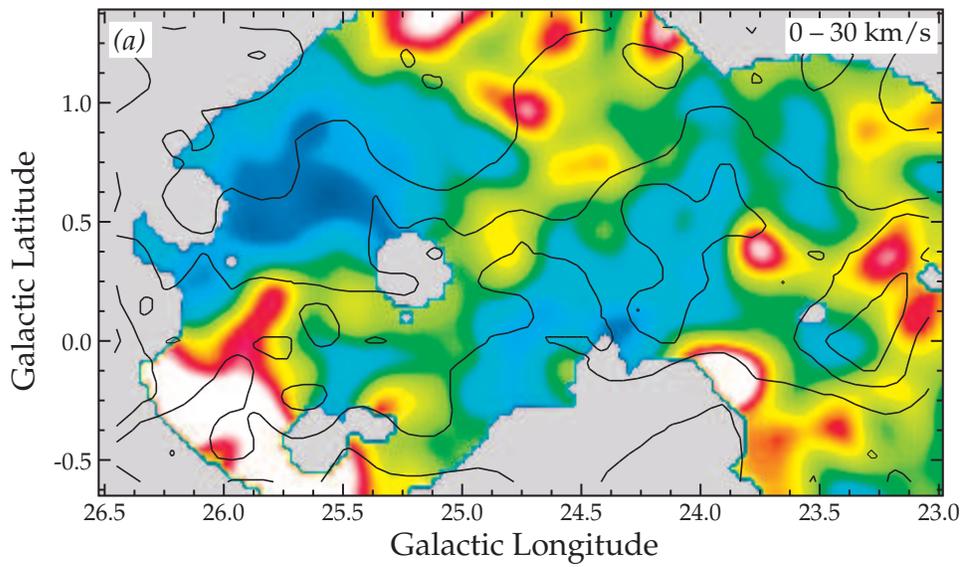
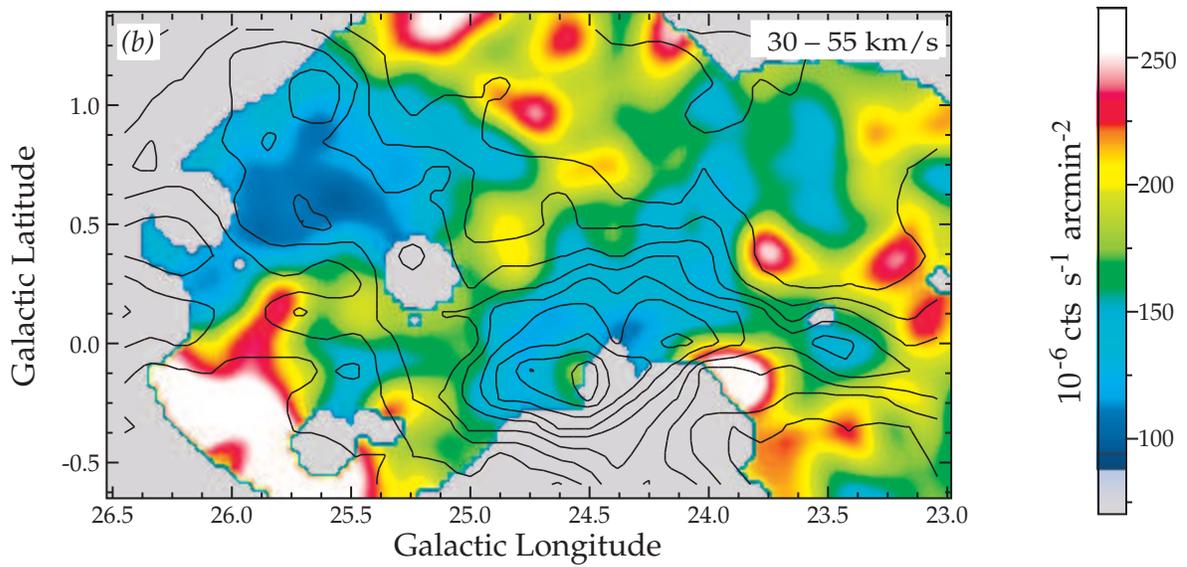
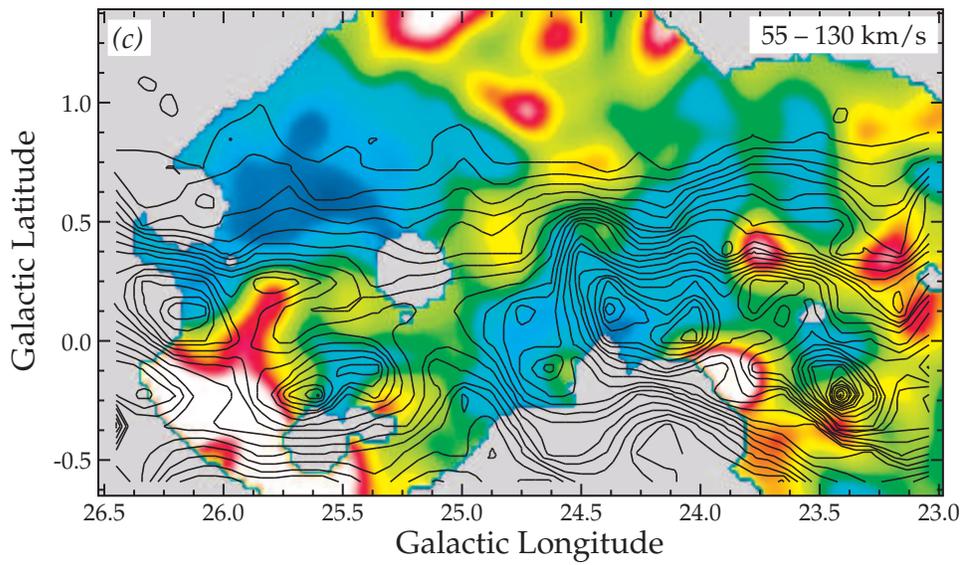
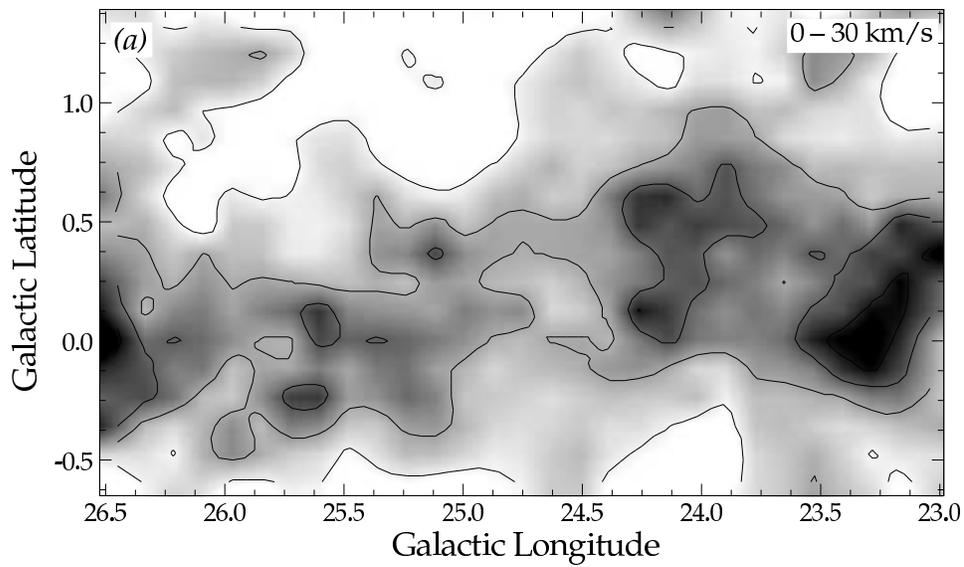
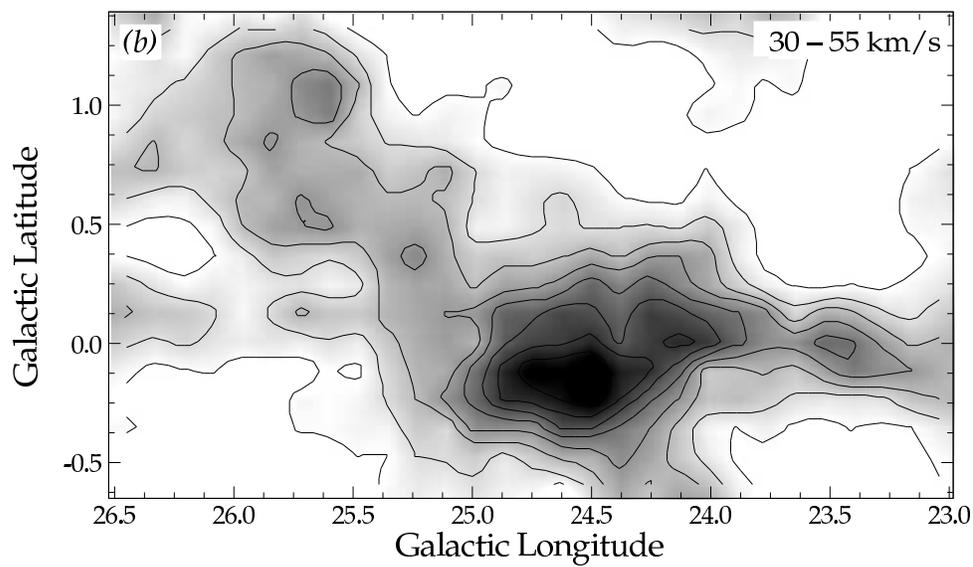
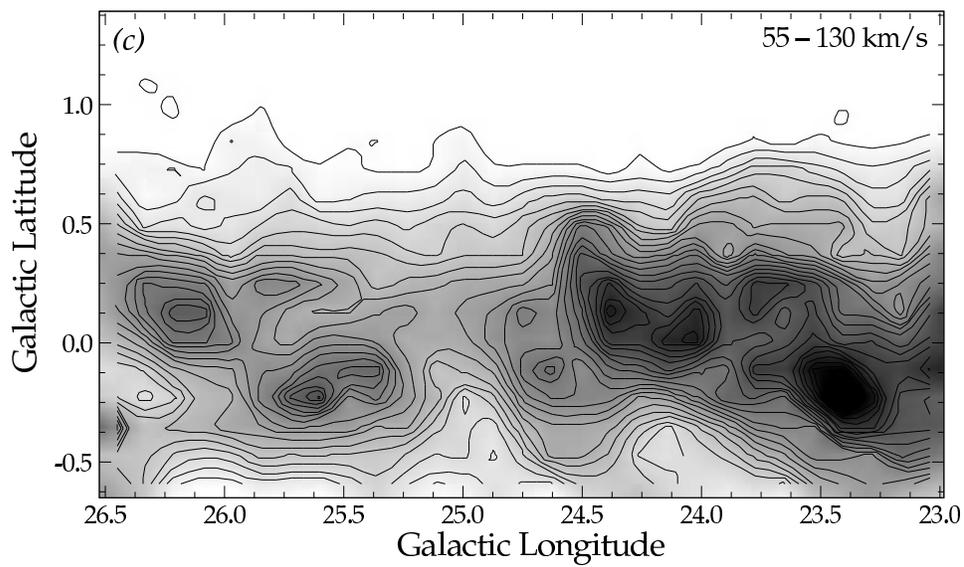
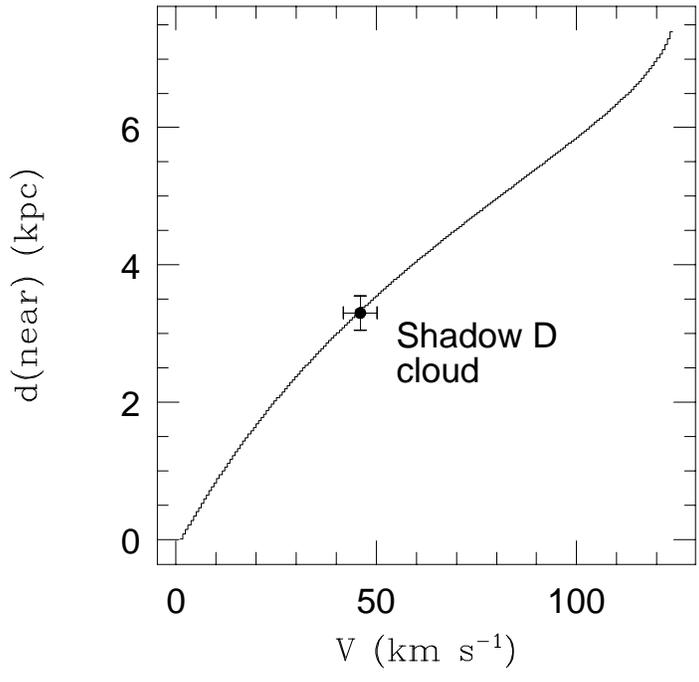
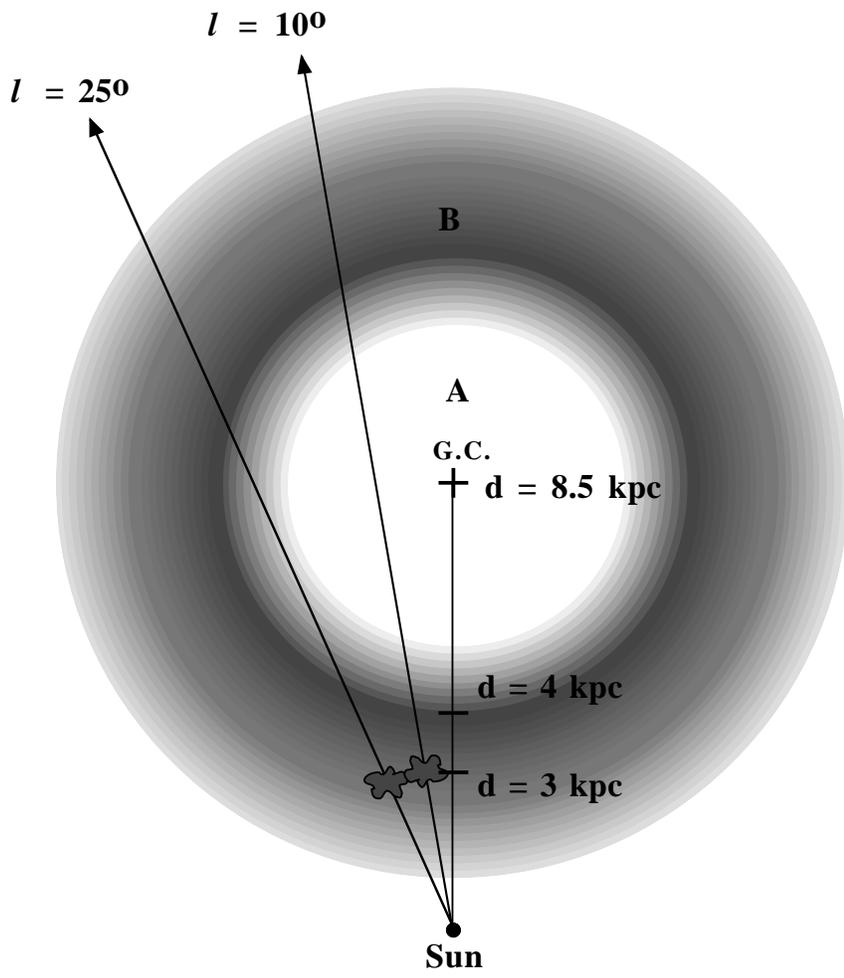
\end{document}